\documentclass[12pt,preprint]{aastex}
\renewcommand{\cite}{\citealp}

\newcommand{\realfigure}[3]{ 
             \hbox{~} \centerline{\includegraphics[width=3.2in]{#1}}
             \figcaption{#2 \label{#3}} \vspace{0.05in}\centerline{}}

\renewcommand{\realfigure}[3]{\placefigure{#3}}

\newcommand{\letter}{{\it Letter}}
\newcommand{\abbrev}[1]{{ #1}}

\newcommand{\allframe}{{\sc allframe}}
\newcommand{\xccdred}{{\sc xccdred}}

\newcommand{\bl}{\phantom{00}}

\newcommand{\distance}{$23.36 \pm 0.17$}  
\newcommand{\Nvarall}{84}
\newcommand{\Nvargood}{43}
\newcommand{\Nrrab}{15}
\newcommand{\Nrrbest}{6}
\newcommand{\Pmean}{$0.613 \pm 0.028$}   
\newcommand{\fehrr}{$-1.92 \pm 0.35$}    
\newcommand{\Vmean}{$24.63 \pm 0.14$}   
\newcommand{\Mvrrnew}{$0.59 \pm 0.03$}
\newcommand{\modlmc}{$18.48 \pm 0.05$}
\newcommand{\Mvrrmygal}{0.50}
\newcommand{\Ebv}{0.25}

\newcommand{\Av}{0.78}

\newcommand{\mygal}{NGC\,6822}
\newcommand{\di}{{dI}}
\newcommand{\rrl}{{RR~Lyrae}}
\newcommand{\gratis}{{GrATiS}}
\newcommand{\acep}{{anomalous Cepheid}}
\newcommand{\aceps}{{\acep s}}

\newcommand{\msol}{$M_\odot$}

\begin{document}

\title{RR~Lyrae and short-period variable stars 
in the dwarf irregular galaxy \mygal\altaffilmark{1}}

\author{
Gisella Clementini,\altaffilmark{2}
Enrico V. Held,\altaffilmark{3}
Lara Baldacci,\altaffilmark{2}$^{,}$\altaffilmark{4}
Luca Rizzi\altaffilmark{3}$^{,}$\altaffilmark{5}
}

\altaffiltext{1}{Based on data collected at E.S.O. La Silla, Chile, 
Proposal~No.~67B--0557}

\altaffiltext{2}{INAF, Osservatorio Astronomico di
Bologna, via Ranzani 1, I-40127 Bologna, Italy;
(gisella, baldacci)@bo.astro.it}

\altaffiltext{3}{INAF, Osservatorio Astronomico di
Padova, vicolo dell'Osservatorio 5, I-35122 Padova, Italy;
(held, rizzi)@pd.astro.it}

\altaffiltext{4}{Dipartimento di Astronomia, Universit\`a di
Bologna, via Ranzani 1, I-40127 Bologna, Italy}

\altaffiltext{5}{Dipartimento di Astronomia, Universit\`a di
Padova, vicolo dell'Osservatorio 2, I-35122 Padova, Italy}

\begin{abstract}
We report the discovery of a large number of short-period variable
stars in the dwarf irregular galaxy \mygal, based on deep time-series
imaging carried out with the ESO Very Large Telescope.  In particular,
we found a modest population of \rrl\ stars tracing the presence of an
old stellar component in \mygal.  Measurements of the average
luminosity of \rrl\ stars provide a new independent estimate of the
distance to this galaxy based on a Pop.~II indicator, $(m-M)_0
=$~\distance.  In addition, our new data show a significant population
of small-amplitude, short-period variable stars filling the instability
strip {starting at luminosities 
only a few tenths of a magnitude brighter than the
\rrl\ stars.}
%
Given the presence of an extended star formation in \mygal, 
{the faint end of this distribution of short-period variable stars} 
is likely to originate
from a population of intermediate-age, metal-poor He-burning stars,
younger and more massive than \rrl\ stars.
\end{abstract}


\section{Introduction}

Extended horizontal branches (HBs) and/or significant
populations of \rrl\ stars have been detected in several dwarf
galaxies 
({NGC\,147: Saha \& Hoessel \cite{saha+hoes87}, Saha et al. \cite{saha+90}; 
NGC\,185: Saha \& Hoessel \cite{saha+hoes90}}; 
{NGC\,205: Saha et al. \cite{saha+92b}}; 
WLM: Rejkuba et al. \cite{rejk+00}; 
Leo\,I: Held et al. \cite{held+00}, \cite{held+01}; 
IC\,1613: Saha et al. \cite{saha+92a}, Dolphin et al. \cite{dolp+01}; 
Leo\,A: Dolphin et al. \cite{dolp+02}, Schulte-Ladbeck et al. \cite{schu+02}; 
And\,VI: Pritzl et al. \cite{prit+02}). 
These observations provide evidence for a first epoch of star
formation which is common to the majority of Local Group dwarf
galaxies, both star forming and presently quiescent. 

The presence of an old stellar population is still an open question in
\mygal, the nearest dwarf irregular (\di) galaxy (and the closest
star-forming galaxy beyond the Magellanic Clouds).  Based on its
luminosity, gas fraction, metallicity, and star formation rate,
\mygal\ appears to be a typical, relatively gently star-forming \di\
galaxy.  Several studies addressed its star formation history by
modeling its color-magnitude diagram (\abbrev{CMD}) (see, e.g.,
Marconi et al. \cite{marc+95}; Gallart et al. \cite{gall+96b}; Wyder
\cite{wyde01}; and references therein).  These studies suggested that
\mygal\ most likely began forming stars 12-15 Gyr ago from
low-metallicity gas. However, the data are not inconsistent with star
formation starting 6-9 Gyr ago from a relatively metal enriched
interstellar medium.

The discovery of \rrl\ variable stars in \mygal\ presented in this
\letter, breaks the uncertainty between these two scenarios by
demonstrating the presence of an old ($>11$ Gyr), metal-poor stellar
population. The average magnitude of the newly discovered \rrl\
variable stars is also used to derive a new estimate of the distance
to this galaxy based on a Pop.~II indicator.

\section{Observations and data reduction}

\subsection{Observations}

Time-series imaging of \mygal\ was obtained on 3 half nights on
Aug.~15, 16, and 20, 2001 using the focal reducer FORS2 on the ESO
VLT/UT3 telescope at Cerro Paranal, Chile. The standard resolution
collimator was used, yielding a $6.8 \times 6.8$ arcmin field-of-view
with a 2048$^2$ pixel Textronix CCD (ESO \#160), read in 4-port mode
with no binning.

The scheduling was devised so as to provide good coverage of the light
curves for variable stars with periods between 0.2 and 0.7 days, with
optimal efficiency for periods around {0.53 and 0.59-0.60 d}.
The observations consisted of $36 \times 900$~s $V$ exposures of
a galaxy field located at $19^h 45^m 13.3^s$, $-14^\circ 45^\prime
55^{\prime\prime}$ (offset from the center of \mygal\ to avoid regions
of active star formation). These data were complemented by 11 $B$ and
1 $I$ exposures of the same duration.
The nights were all photometric and the seeing varied from {0.5}
to 1.0 arcsec, with a median value around 0.7 arcsec.  
%

\subsection{Photometry}

The FORS2 images were reduced in a standard way using the \xccdred\
IRAF\footnote{IRAF is distributed by the National Optical Astronomical 
Observatories, which are operated by the Association of Universities for
Research in Astronomy, Inc., under cooperative agreement with the 
National Science Foundation} package.  Photometry of
all bias-subtracted, flat-fielded images was obtained using \allframe\
(Stetson \cite{stet94}). 
Since our standard stars were mostly saturated even with the shortest
exposure times allowed by the instrument software, the instrumental
$b$,$v$, and $i$ magnitudes were calibrated using the color equations
available from the FORS Web page. The photometric zero points were
then set by comparison with secondary standard stars established in
the field using Wide Field Imager observations at the 2.2m ESO/MPI
telescope, accurately calibrated onto the Landolt (\cite{land92})
system (L. Rizzi et al. 2003, in prep.). The estimated uncertainty of
the zero point calibration is 0.04 mag in both $B$ and $V$.


\subsection{Variable star identification and period search}

Candidate variable stars were identified using ISIS2.1 (Alard
\cite{alar00}), a package specifically designed for the detection of
variable stars in crowded fields.  ISIS 
is based on an optimal image subtraction technique. The 
variable star detection procedure consists
of the following steps: (i) geometric alignment of the image series
and re-mapping of each image to the same grid; (ii) construction of a
reference image obtained by stacking a subset of images; (iii)
subtraction of each individual frame from the reference image
convolved with a suitable kernel to match seeing variations and
geometrical distortions on the individual images; (iv) photometry of
the variable objects on the difference images, in terms of
differential fluxes.
%
The search performed by ISIS on this
image yielded a master list of 448 candidate variable stars. 
Candidate variable stars
were then counter-identified on
the \allframe\ $B$,$V$ master catalogue, where about 2/3 of the
objects flagged by ISIS as variable sources were found.

All stars in the candidate master list were then analyzed to confirm
their variable star nature and derive light curves. 
For the analysis, we combined the
time series obtained from \allframe\ photometry (whenever available)
and from ISIS differential flux measurements.
In this \letter,
given our primary interest in \rrl\ variable stars, only stars with
mean magnitude fainter than $V = 23$ are discussed. A full census 
of the
variable stars {identified} 
in \mygal\ will be given in a forthcoming paper
(E. V. Held et al. 2003, in prep.). 

For the period search we used the program \gratis\ (Graphical Analyzer
of Time Series: see Clementini et al. \cite{clem+00} for details),
which was run on the ISIS differential flux data converted to
magnitude scale.  Whenever multiple periods appeared equally
compatible with the data, the star was omitted from the present ``best
sample''.
The light curves were finally calibrated to the standard photometric
system by modeling the color variation along the cycle.  The sampling
of the $B$ data points, although sparser than that of the $V$
observations, is sufficient to constrain the model color curves quite
well.

\section{Results}

\subsection{Light curves}

Among the \Nvarall\ candidates in the faint sample considered in this
\letter, \Nvargood\
have well sampled light curves in both $V$ and $B$, and a scatter about the
light curve model not larger than 0.1 mag.  This restricted sample 
was found to 
includes 17 \rrl, 20 brighter short-period
variable stars with very similar period distributions, and 6 further
variables 
among which 
some 
binary systems and some possible $\beta$
Cephei stars.




For the classification of 
{the faint sample of variable stars} 
we took into 
account the mean luminosity and color of the stars, and the period,
amplitude, and shape of the light curve.
Based on 
{these characteristics} our sample was
found to include \rrl\ stars and brighter stars pulsating both in the
fundamental mode (the vast majority) and in the first overtone.
%
The mean period of the \Nrrab\ $ab$-type \rrl\ variable
stars with the best defined light curves is ${\langle P_{ab}\rangle}
=$~\Pmean, where the uncertainty is the stardard error on the mean.

\realfigure{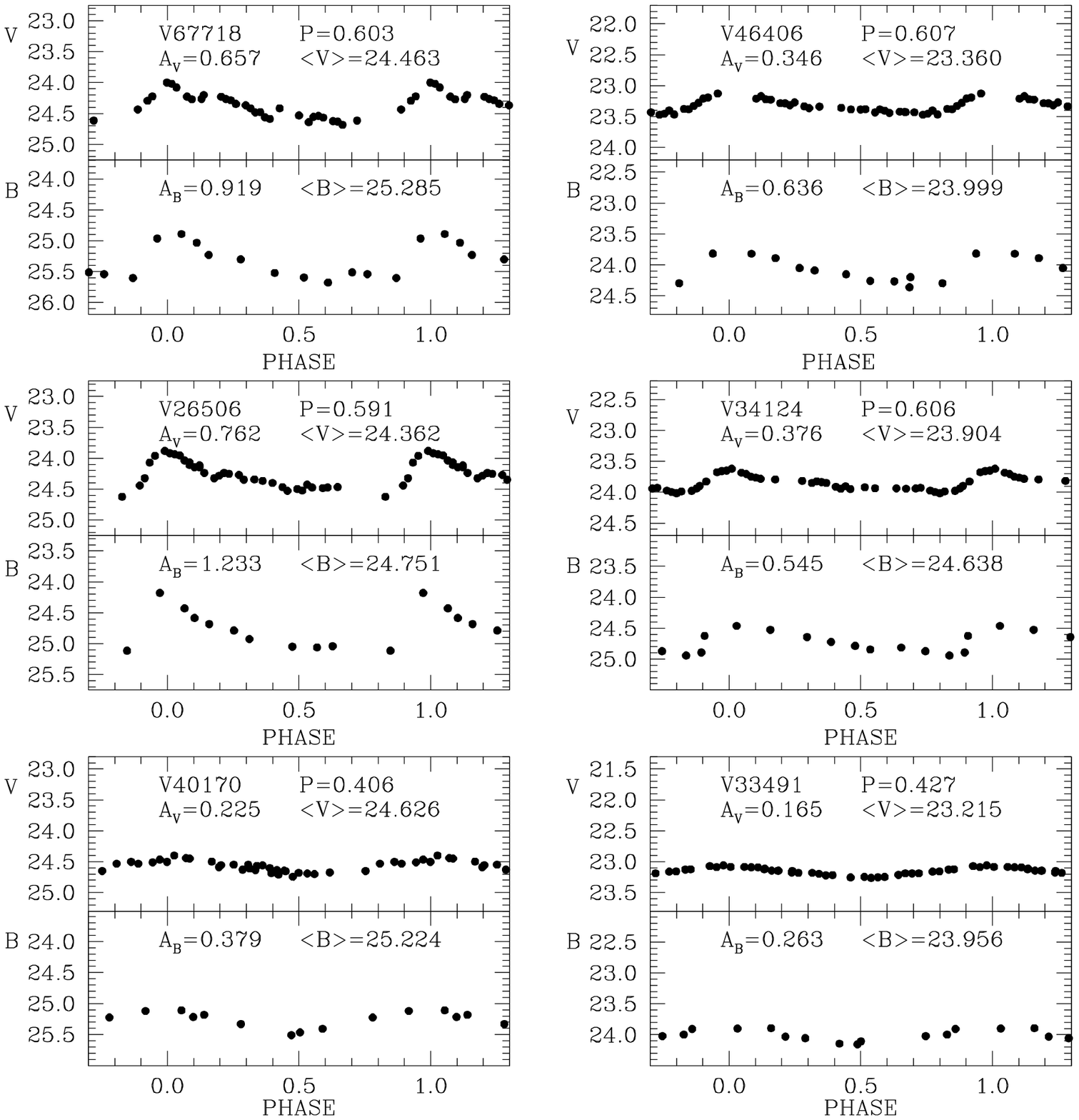} 
{$B$ and $V$ light curves
of short-period variable stars in \mygal.  The {\em left} panels are
\rrl\ stars, while light curves on the {\em right} panels refer to
brighter stars. A constant magnitude interval is used in all panels to
facilitate the comparison.}  
{f_lcurves}

\begin{figure}\plotone{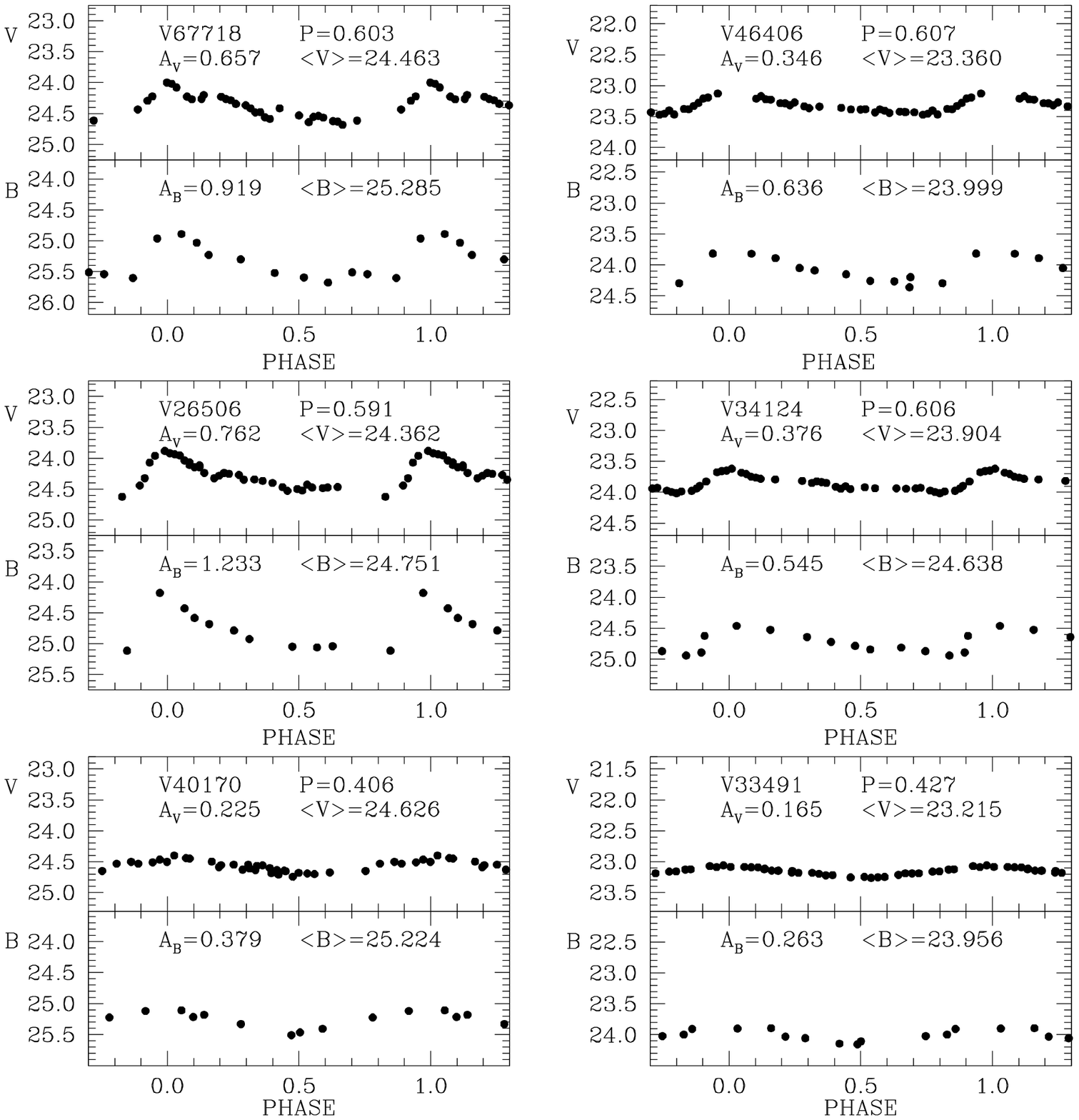} 
\caption{$B$ and $V$ light curves
of short-period variable stars in \mygal.  The {\em left} panels are
\rrl\ stars, while light curves on the {\em right} panels refer to
brighter stars. A constant magnitude interval is used in all panels to
facilitate the comparison.  
\label{f_lcurves}
}\end{figure}

Figure~\ref{f_lcurves} shows some examples of the $V$ and $B$ light
curves of variable stars discovered in \mygal.  Labels in each panel
give the identifier, period, amplitude and intensity-averaged magnitude
($V$ and $B$) of each variable star.  The left panels show typical
light curves of \rrl\ variable stars, both fundamental mode (top and
middle) and first overtone pulsators ($c$-type \rrl, bottom panel).
The light curves on the right panels refer to some of the numerous
pulsating stars
that have been discovered in \mygal. These variables are
brighter than the
\rrl\ stars and have on average smaller amplitudes at a given period.
Because of their characteristics these stars appear related to
both the short-period Population I Cepheids found in dwarf irregular
galaxies (Dolphin et al. \cite{dolp+01}, \cite{dolp+02},
\cite{dolp+03}; Sharpee et al. \cite{shar+02}) and the \aceps\
population in dwarf spheroidal galaxies (e.g., Nemec, Nemec, \& Lutz
\cite{neme+94}; Bersier \& Wood \cite{bers+wood02}).
The first two light curves (top and middle right panels) have
asymmetric shape similar to $ab$-type \rrl\ stars, however they have
lower amplitudes by about a factor of two, and 
{the mean magnitudes, calculated from the average of the light curves,} 
are about 
{0.4} to 1 mag brighter.  The curve on the last right panel, 
has a sinusoidal shape
similar to $c$-type pulsators, yet its mean magnitude, about 1.4 mag
brighter than the \rrl\ stars, clearly qualifies this star as a
Cepheid.

\subsection{Variable stars in the color-magnitude diagram}

\begin{figure}\plotone{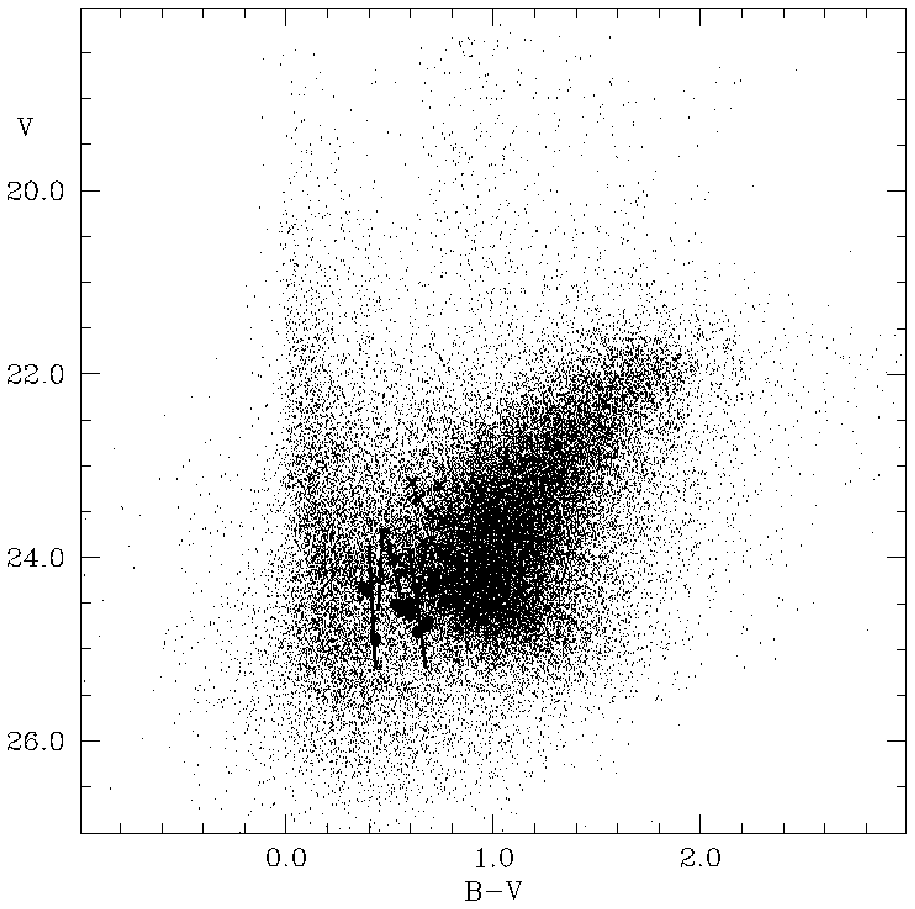} 
\caption{The calibrated color-magnitude
diagram of \mygal\ showing the location of the newly discovered faint
pulsating variable stars.  
The {\em filled circles} are
stars classified \rrl, {\em crosses} represent 
{small-amplitude}, short-period Cepheids.
The {\em solid
lines} show the edges of the instability strip in the globular cluster
M\,3 (Corwin \& Carney \cite{corw+carn01}), while the {\em dashed
lines} are the limits for pulsational instability for 1.5 \msol\
models (Bono et al. \cite{bono+97}), converted to the observational
plane using the model atmospheres of Castelli, Gratton, \& Kurucz
(\cite{cast+97}).
\label{f_diag}
}\end{figure}

The location of the faint variable star sample in the color-magnitude
diagram of \mygal\ is shown in Fig.~\ref{f_diag}.  Only 
pulsating variable stars with complete light curves and good
photometry have been plotted, using their intensity-averaged
magnitudes and colors.  Different symbols have been used for \rrl\
stars and more luminous short-period variable stars.  Since no HB
could be identified even from HST data (Wyder \cite{wyde01}), our
detection of \rrl\ stars provides {\it the first {indisputable}
evidence for the presence of an old population in \mygal}. In fact,
while the HB itself is completely hidden by the overwhelming number of
young and intermediate-age stars that populate the same region of the
diagram, the position of the \rrl\ variable stars agrees well with the
HB level and the instability strip in the Galactic cluster M\,3
(Corwin \& Carney \cite{corw+carn01}), shifted to the distance and
reddening of \mygal.  We adopted the foreground reddening $E_{B-V} =
$~\Ebv\ from the maps of Schlegel, Finkbeiner, \& Davis
(\cite{schl+98}), with no account for internal reddening. The scatter
of the data on the red side of the instability strip is probably due
to incomplete phase coverage of the light curves.

In addition to \rrl\ stars, a continuous distribution of short-period
variable stars is shown in Fig.~\ref{f_diag}, filling the region of
the diagram above the HB, beginning just about 0.3 -- 0.5 mag brighter
than the average luminosity level of the \rrl\ stars, and with 
slightly redder mean colors. The
observed mean colors and luminosities are not inconsistent with the
limits for pulsational instability predicted by theoretical models for
the 1.5 \msol\ models of \aceps\ (Bono et al. \cite{bono+97}).

\realfigure{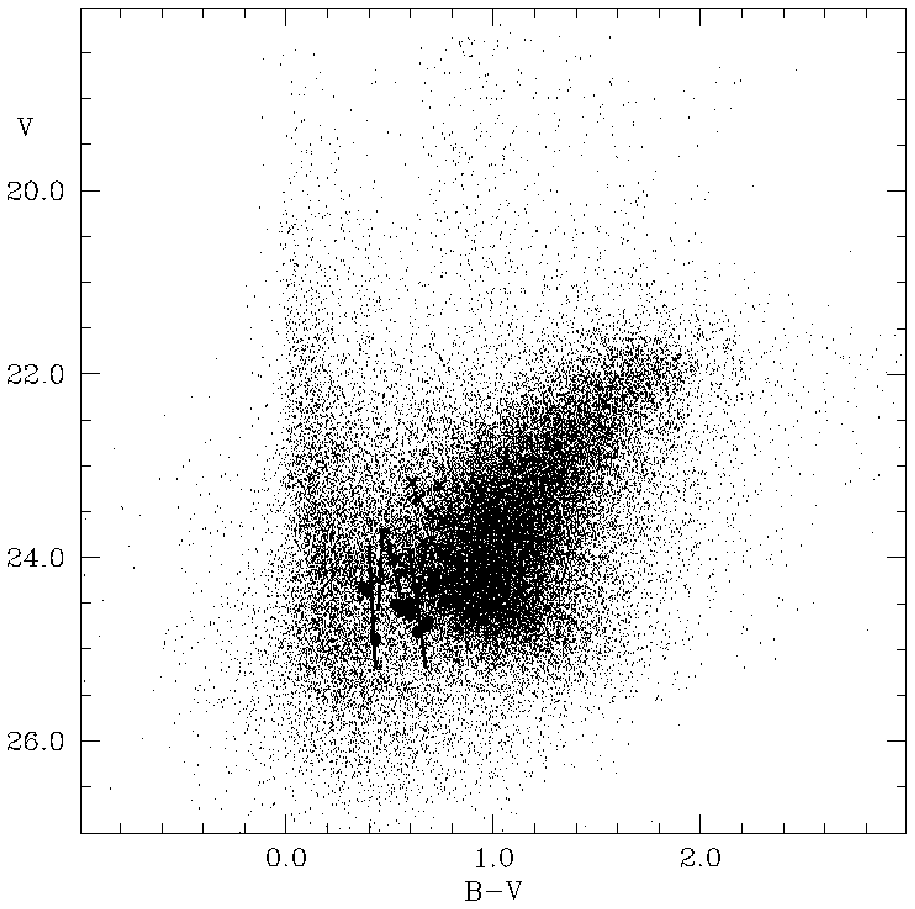} {The calibrated color-magnitude
diagram of \mygal\ showing the location of the newly discovered faint
pulsating variable stars.  
The {\em filled circles} are
stars classified \rrl, {\em crosses} represent 
short-period Cepheids.
The {\em solid
lines} show the edges of the instability strip in the globular cluster
M\,3 (Corwin \& Carney \cite{corw+carn01}), while the {\em dashed
lines} are the limits for pulsational instability for 1.5 \msol\
models (Bono et al. \cite{bono+97}), converted to the observational
plane using the model atmospheres of Castelli, Gratton, \& Kurucz
(\cite{cast+97}).}{f_diag}

\section{Analysis}

\subsection{The period-amplitude relation}

\begin{figure}\plotone{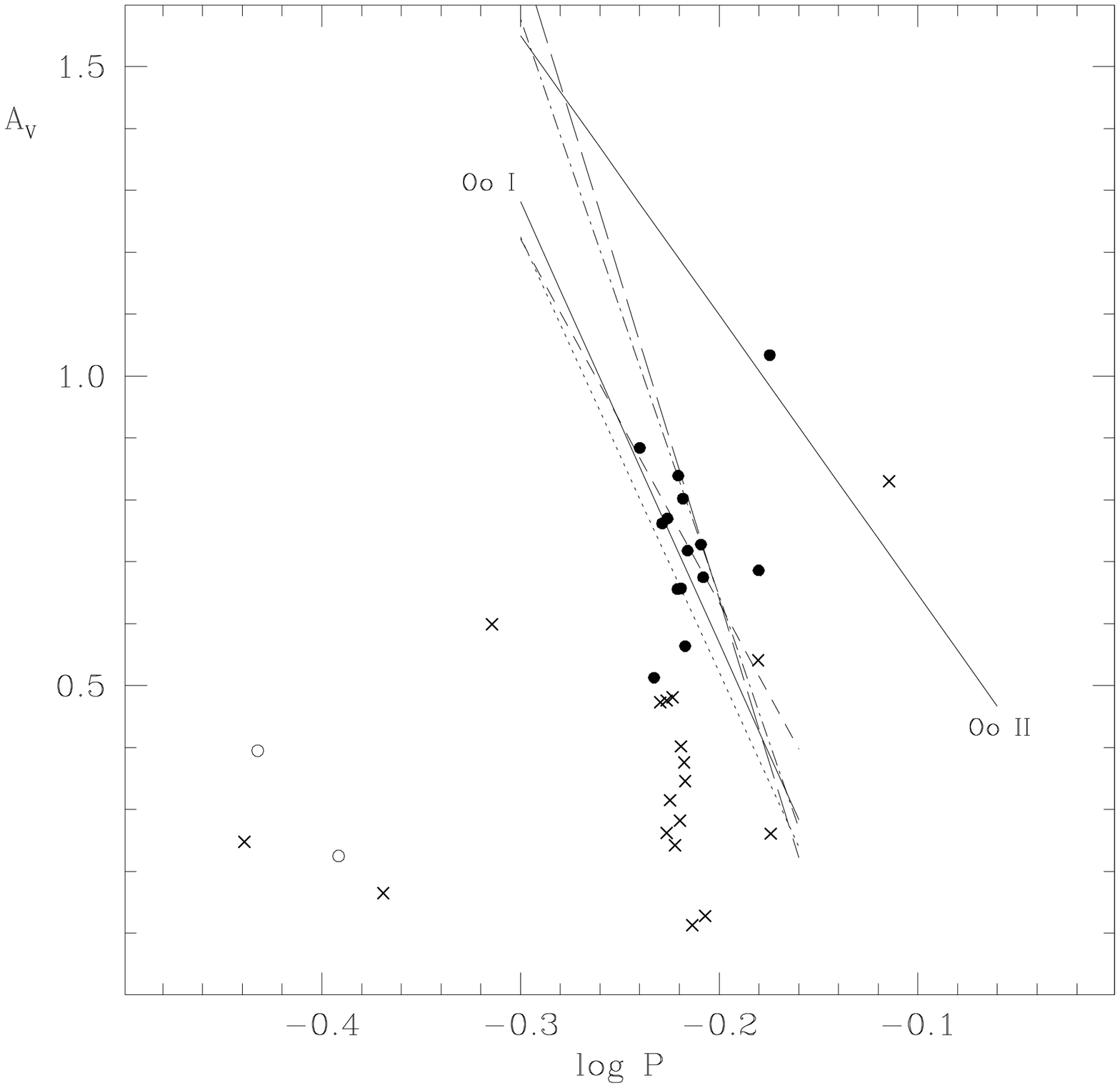}
\caption{Period-amplitude relation for \rrl\ and other short-period variable
stars with complete light curves 
in \mygal. Periods are in days.  {\em Filled circles} and {\em
open circles} are $ab$-type and $c$-type \rrl\ stars in \mygal,
respectively.  The {\em crosses} represent the sequence of
{small-amplitude, short-period} Cepheids. 
The lines show the $A_V - \log P$ relations for the
$ab$-type variable stars in Galactic globular clusters ({\em
continuous lines}, from Clement \& Rowe \cite{cle+00}) and dwarf
spheroidal galaxies (And VI: {\em dots}; Sculptor: {\em short dash};
Leo II: {\em long dash}; Draco: {\em dot-short dash}; from Pritzl et
al. \cite{prit+02}). 
\label{f_pav}
}\end{figure}

Figure~\ref{f_pav} shows the period-amplitude relation for the
{faint} variable stars detected in \mygal\ 
{(namely \rrl\ and short-period variables fainter than $V$=23)} 
for which complete light curves have been obtained.

In several dwarf spheroidal galaxies of the Local Group (Leo~II,
Leo~I, Draco), \rrl\ stars span a wide range in pulsational
properties, suggesting a classification intemediate between the
Oosterhoff~I and II groups 
(Oosterhoff \cite{oo}) 
of Galactic globular clusters (Siegel \&
Majewski \cite{sieg+maje00}; Held et al. \cite{held+01}; Kinemuchi et
al. \cite{kine+02}).  In \mygal, the period and amplitude
distributions seem more biased towards an Oosterhoff~I type,
similarly to \rrl\ stars in Sculptor (Kaluzny et al. \cite{kalu+95}),
the LMC (Alcock et al. \cite{alco+96}) and And~VI (Pritzl et
al. \cite{prit+02}), although a composite distribution is not ruled
out by our data.
Indeed, the \rrl\ sample in Fig.~\ref{f_pav} is subject to several
biases both because \rrl\ stars are found close to limiting magnitude
of our photometry, and because of the short baseline of our
observations.
%
%
%

\realfigure{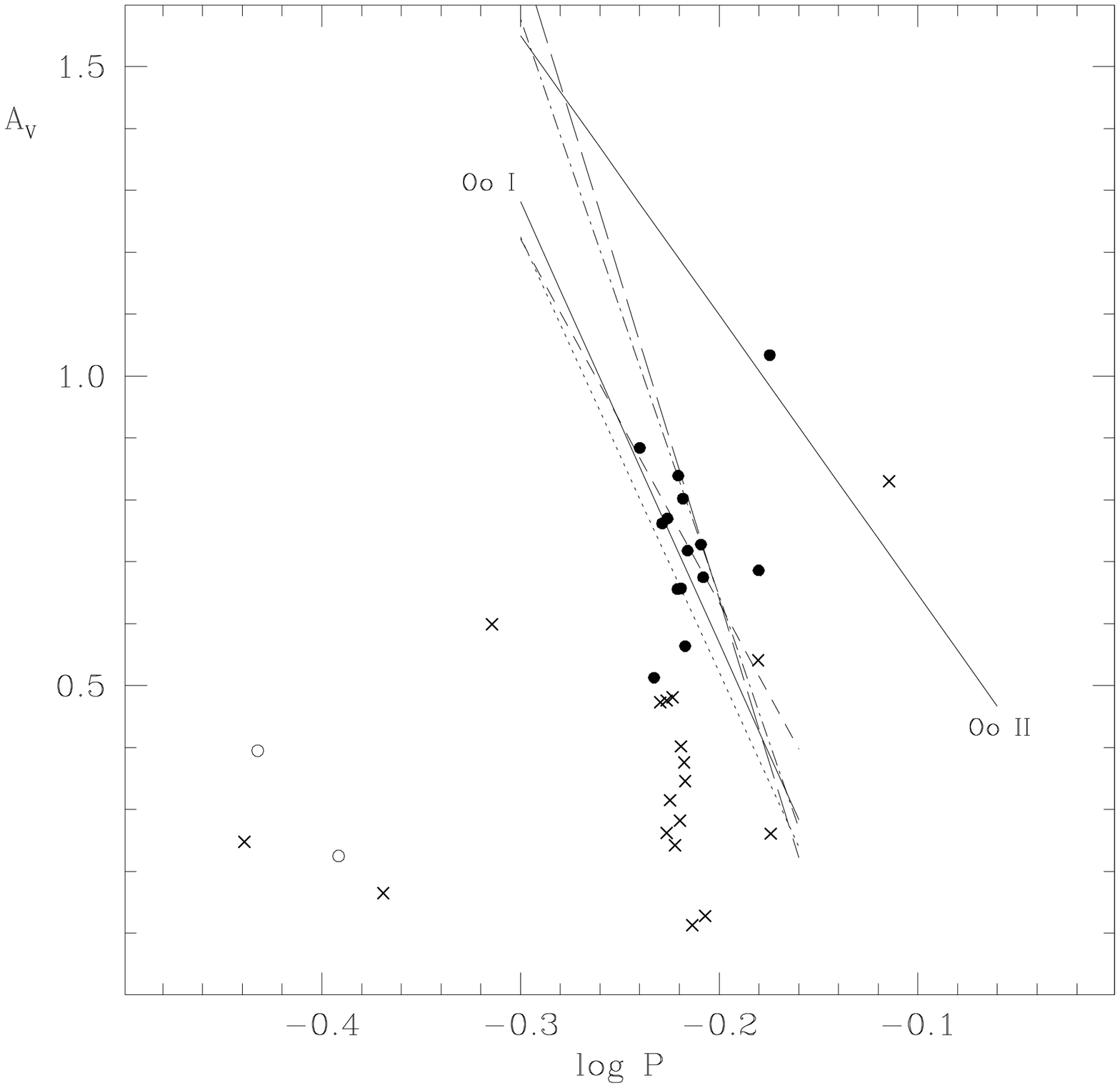}
{Period-amplitude relation for \rrl\ and other short-period variable
stars with complete light curves 
in \mygal. Periods are in days.  {\em Filled circles} and {\em
open circles} are $ab$-type and $c$-type \rrl\ stars in \mygal,
respectively.  The {\em crosses} represent the sequence of
candidate \aceps.  The lines show the $A_V - \log P$ relations for the
$ab$-type variable stars in Galactic globular clusters ({\em
continuous lines}, from Clement \& Rowe \cite{cle+00}) and dwarf
spheroidal galaxies (And VI: {\em dots}; Sculptor: {\em short dash};
Leo II: {\em long dash}; Draco: {\em dot-short dash}; from Pritzl et
al. \cite{prit+02}). 
}{f_pav}

{What is the nature of the small-amplitude, short-period variable
stars found in \mygal\ 
at luminosities only a few tenths of a
magnitude brighter than the \rrl\ stars ?  To the best of our knowledge,
{Cepheids} with such low luminosities (LLC), small amplitudes, and
short periods (see Fig.~\ref{f_pav}) have not been found before in any
dwarf irregular galaxy, although 
they are consistent with 
the predictions of theoretical pulsational models 
(Fiorentino et al. \cite{fior+03}; F. Caputo et al. 2003, in prep.).
Given the extended star-formation history of \mygal, these LLC
are likely to originate from a population of intermediate-age,
metal-poor He-burning stars
younger and more massive than the \rrl\ stars. The brighter variable
stars appear more similar to the short-period Cepheids found in 
other dwarf
irregular galaxies (e.g., Dolphin et al. \cite{dolp+03}).
A full discussion of the characteristics of this population of
short-period variable stars, in particular the LLC, is postponed to a
forthcoming paper (E.~V.~Held et al. 2003, in prep.).

}

\subsection{Metallicity of the old stellar component in \mygal}



The mean metallicity of \rrl\ stars in \mygal\
can be estimated using the relation between the mean period and
metallicity of $ab$-type \rrl\ stars derived by Sandage
(\cite{sand93}) for variable stars in Galactic globular clusters
(GGCs),
$\log{\langle P_{ab} \rangle} = -0.092~\mbox{[Fe/H]} -0.389$.
From the average period of the $ab$-type \rrl\ stars with well-defined
light curves, we derived [Fe/H]$ = $\fehrr, where the error 
{includes} the
standard deviation of the measurements, obtained by error propagation
from the scatter of the period distribution, and uncertainties
due to detection efficiency and possible systematics if 
\rrl\ in \mygal\ are younger than those in GGCs.
{
This value is in remarkable agreement with the low metal abundance
measured for an old globular cluster in \mygal\ (Cohen \& Blakeslee
\cite{cohen+blak98}; Chandar, Bianchi, \& Ford \cite{chand+00}), and
represents the first estimate of the metallicity of the old ``field''
stellar population in \mygal. This value is only slightly more
metal-poor than the metal abundance inferred for RGB stars ([Fe/H]$=
-1.5 \pm 0.3$: Gallart, Aparicio, \& Vilchez \cite{gall+96a}; $-2 <
\mbox{[Fe/H]} < -0.5$: Tolstoy et al.  \cite{tols+01}), which is
consistent with a plausible chemical enrichment law.  

%
%
}


\subsection{The distance to \mygal}

The mean magnitude of \rrl\ variable stars in \mygal\ provides an
independent distance estimate based on Pop.~II stars and offers us the
possibility to compare the distance scales based on Pop.~I and Pop.~II
indicators. We calculated the mean apparent magnitude of the \Nrrbest\
\rrl\ variable stars fainter than $V=24.2$ with well-defined light
curves and located inside the instability strip.  The mean value is
\Vmean, where the error is the standard deviation of the data.

For the absolute magnitude of \rrl\ variable stars, we follow here the
recent re-evalution of Globular Cluster distances by Cacciari \&
Clementini (\cite{cacc+clem03}), suggesting for the absolute magnitude
of \rrl\ stars $M_V(\mbox{RR}) = $~\Mvrrnew\ at [Fe/H]$=-1.5$,
corresponding to a distance modulus for the LMC $(m-M)_0 = $~\modlmc
.
To correct the absolute magnitude to the metallicity of \mygal, we
adopt the mean metallicity of the old population derived from the mean
period of $ab$-type \rrl\ stars, along with a luminosity-metallicity
dependence of 0.22 mag per metallicity dex (as the average of Cacciari
\& Clementini \cite{cacc+clem03} and Clementini et
al. \cite{clem+03}).  These assumptions yield $M_V(\mbox{RR}) =
$~\Mvrrmygal\ for \mygal\ and a distance modulus $(m-M)_0
=$~\distance. The error includes both the systematic error on the
photometry zero point and the uncertainty of the mean apparent and
absolute magnitude of \mygal\ \rrl\ stars, 
%
but does not take into account the contribution of possible
internal differential reddening. 


Table~\ref{t_dist} compares this new estimate of the distance to
\mygal\ with distance moduli previously derived using different
methods.  The new estimate from the \rrl\ stars agrees well with the
distance moduli based on the tip of the red giant branch (Gallart et
al. \cite{gall+96a}) and infrared photometry of Cepheids (Mc~Alary et
al. \cite{mcal+83}), while there is a tendency for the measurements
based on the Cepheid period-luminosity relation to yield a somewhat
longer distance, 
{a difference that might be ascribed to variable reddening (see, e.g.,
Venn et al. \cite{venn+01}, and references therein).  A similar
tendency was noted in other galaxies, most recently by
Clementini et al. (\cite{clem+03}) and Dolphin et
al. (\cite{dolp+03}).}

\acknowledgments 
It is a pleasure to thank M. Marconi for fruitful discussions on
stellar pulsation models, 
 and the referee for helpful remarks.  We
acknowledge support from the National Projects 
COFIN2001028897 and COFIN2002028935.



\clearpage


{
\footnotesize
\tablecaption{Distance determinations to \mygal \label{t_dist}}
\begin{deluxetable}{lccll}
\tablewidth{0pt}
\tablehead{
\colhead{$(m-M)_0$} &
\colhead{$E_{B-V}$} &
\colhead{$A_V$} &
\colhead{Method} &
\colhead{Ref.}
}
\startdata
$ 23.40 \pm 0.11$ &   0.36   &   1.19 &\bl Ceph &     Mc~Alary et al. (1983) \\
$ 23.62         $ &   0.28   &   0.92 &\bl Ceph &     Lee et al. (1993)      \\
$ 23.46         $ &   0.28   &   0.92 &\bl TRGB &     Lee et al. (1993)      \\
$ 23.49 \pm 0.08$ &   0.24   &   0.80 &\bl Ceph &     Gallart et al. (1996a)  \\
$ 23.40 \pm 0.10$ &   0.24   &   0.80 &\bl TRGB &     Gallart et al. (1996a)  \\
  \distance       &   \Ebv   &   \Av  &\bl \rrl &     this work            \\
\enddata
\end{deluxetable}
}

\clearpage

\begin{thebibliography}{}

\bibitem[2000]{alar00}
  Alard, C.\ 2000, \aaps, 144, 363 

\bibitem[1996]{alco+96}
  Alcock, C.~et al.\ 1996, \aj, 111, 1146 

\bibitem[2001]{bers+wood02}  
  Bersier, D.,~\& Wood, P.~R.\ 2002, \aj, 123, 840 

\bibitem[1997]{bono+97}
  Bono, G., Caputo, F., Santolamazza, P., Cassisi, S., \& Piersimoni, A.\ 
  1997, \aj, 113, 2209

\bibitem[2003]{cacc+clem03}
  Cacciari, C., \& Clementini, G. 2003, in Stellar Candles, Lecture
  Notes in Physics (Berlin:Springer), in press (astro-ph/0301550)

%

\bibitem[1997]{cast+97}
  Castelli, F., Gratton, R.~G., \& Kurucz, R.~L.\ 1997, \aap, 318, 841

\bibitem[2000]{chand+00}
  Chandar, R., Bianchi, L., \& Ford, H.~C.\ 2000, \aj, 120, 3088

\bibitem[2000]{clem+00}
  Clementini, G.~et al.\ 2000, \aj, 120, 2054 

\bibitem[2003]{clem+03} 
  Clementini, G., Gratton, R.~G., Bragaglia, A.,
  Carretta, E., Di Fabrizio, L., \& Maio, M.\ 2003, \aj, 125, 1309
   
\bibitem[2000]{cle+00}
  Clement, C.~M., \& Rowe, J.\ 2000, \aj, 120, 2579

\bibitem[1998]{cohen+blak98}
  Cohen, J.~G., \& Blakeslee, J.~P.\ 1998, \aj, 115, 2356

\bibitem[2001]{corw+carn01}
  Corwin, T.~M., \& Carney, B.~W.\ 2001, \aj, 122, 3183

\bibitem[2001]{dolp+01}
  Dolphin, A.~E.~et al.\ 2001, \apj, 550, 554     

\bibitem[2002]{dolp+02}
  Dolphin, A.~E.~et al.\ 2002, \aj, 123, 3154    

\bibitem[2003]{dolp+03}
  Dolphin, A.~E.~et al.\ 2003, AJ, 125, 1261   

\bibitem[2003]{fior+03} 
  Fiorentino, G., Caputo, F., \& Marconi, M. 2003, in Stars in
  Galaxies, eds. M. Bellazzini, A. Buzzoni, S. Cassisi, MSAIt, in
  press

\bibitem[1996a]{gall+96a}  
  Gallart, C., Aparicio, A., \& 
  Vilchez, J.~M.\ 1996, \aj, 112, 1928 

\bibitem[1996b]{gall+96b}
  Gallart, C., Aparicio, A., Bertelli, G., \& Chiosi, C.\ 1996, \aj, 112, 1950 


\bibitem[2000]{held+00}
  Held, E.~V., Saviane, I., Momany, Y., \& Carraro, G.\ 2000, 
  \apjl, 530, L85

\bibitem[2001]{held+01}
  Held, E.~V., Clementini, G., Rizzi, L., Momany, Y., Saviane, I., \&
  Di Fabrizio, L.\ 2001, \apjl, 562, L39

\bibitem[1995]{kalu+95}
  Kaluzny, J., Kubiak, M., Szymanski, M., Udalski, A., Krzeminski, W.,
  \& Mateo, M.\ 1995, \aaps, 112, 407

\bibitem[2002]{kine+02} 
  Kinemuchi, K., Smith, H.~A., Lacluyz{\' e}, A.~P., Clark, C.~L.,
  Harris, H.~C., Silbermann, N., \& Snyder, L.~A.\ 2002, ASP
  Conf.~Ser.~259: IAU Colloq.~185: Radial and Nonradial Pulsationsn as
  Probes of Stellar Physics (San Francisco:ASP), 130

\bibitem[1992]{land92}
  Landolt, A. U. 1992, AJ, 104, 340

\bibitem[1993]{lee+93}
  Lee, M.~G., Freedman, W.~L., \& Madore, B.~F.\ 1993, \apj, 417, 553

\bibitem[1995]{marc+95}
  Marconi, G., Tosi, M., Greggio, L., \& Focardi, P.\ 1995, \aj, 109, 173



\bibitem[1983]{mcal+83}
  McAlary, C.~W., Madore, B.~F., McGonegal, R., McLaren, R.~A., \&
  Welch, D.~L.\ 1983, \apj, 273, 539

\bibitem[1994]{neme+94}
  Nemec, J. M., Nemec, A. F. L., \& Lutz T. E. 1994, AJ, 108, 222

\bibitem[1939]{oo}
  Oosterhoff, P. Th. 1939, Observatory, 62, 104


\bibitem[2002]{prit+02}
  Pritzl, B.~J., Armandroff, T.~E., Jacoby, G.~H., \& Da Costa, G.~S.\
  2002, \aj, 124, 1464

\bibitem[2000]{rejk+00}
  Rejkuba, M., Minniti, D., Gregg, M.~D., Zijlstra, A.~A., Alonso,
  M.~V., \& Goudfrooij, P.\ 2000, \aj, 120, 801

\bibitem[1992a]{saha+92a}
  Saha, A., Freedman, W.~L., Hoessel, J.~G., \& Mossman, A.~E.\ 1992a,
  \aj, 104, 1072

\bibitem[1987]{saha+hoes87}
  Saha, A., \& Hoessel, J.~G.\ 1990,
  \aj, 94, 1556

\bibitem[1990]{saha+hoes90}
  Saha, A., \& Hoessel, J.~G.\ 1990,
  \aj, 99, 97

\bibitem[1990]{saha+90}
  Saha, A., Hoessel, J.~G., \& Mossman, A.~E.\ 1990,
  \aj, 100, 108

\bibitem[1992b]{saha+92b}
  Saha, A., Hoessel, J.~G., \& Krist, J.\ 1992b,
  \aj, 103, 84

\bibitem[1993]{sand93}
  Sandage, A.\ 1993, \aj, 106, 687 

\bibitem[1998]{schl+98}
  Schlegel, D.~J., Finkbeiner, D.~P., \& Davis, M.\ 1998, \apj, 500, 525 

\bibitem[2002]{schu+02}
  Schulte-Ladbeck, R.~E., Hopp, U., Drozdovsky, I.~O., Greggio, L., \&
  Crone, M.~M.\ 2002, \aj, 124, 896

\bibitem[2002]{shar+02}
  Sharpee, B., Stark, M., Pritzl, B., Smith, H., Silbermann, N.,
  Wilhelm, R., \& Walker, A.\ 2002, \aj, 123, 3216

\bibitem[2000]{sieg+maje00}
  Siegel, M.~H., \& Majewski, S.~R.\ 2000, \aj, 120, 284

\bibitem[1994]{stet94}
  Stetson, P. B. 1994, PASP, 106, 250

\bibitem[2001]{tols+01} 
  Tolstoy, E., Irwin, M.~J., Cole, A.~A., Pasquini, L., Gilmozzi, R., \&
  Gallagher, J.~S.\ 2001, \mnras, 327, 918


\bibitem[2001]{venn+01}
  Venn, K.~A.~et al.\ 2001, \apj, 547, 765 


\bibitem[2001]{wyde01}
  Wyder, T.~K.\ 2001, \aj, 122, 2490

\end{thebibliography}
\end{document}